\documentclass[article]{IEEEtran}
\IEEEoverridecommandlockouts
\usepackage{spconf}

\usepackage[numbers,sort&compress,square]{natbib}
\usepackage{amsmath,amssymb,amsfonts}
\usepackage{bm}
\usepackage{algorithmic}

\usepackage{cleveref}

\usepackage{graphicx}
\usepackage{caption}
\usepackage{subcaption}

\usepackage{textcomp}
\usepackage{xcolor}
\usepackage{url}
\usepackage{balance}

\def\BibTeX{{\rm B\kern-.05em{\sc i\kern-.025em b}\kern-.08em
    T\kern-.1667em\lower.7ex\hbox{E}\kern-.125emX}}
\newcommand{\tsne}{$t$-SNE }
\newcommand{\etal}{\textit{et al.} }
\renewcommand\thesection{\arabic{section}}
\renewcommand\thesubsection{\thesection.\arabic{subsection}}

\title{The Effects of Signal-to-Noise Ratio on Generative Adversarial Networks Applied to Marine Bioacoustic Data\\
\thanks{\tiny{This work was supported by the Engineering and Physical Sciences Research Council, Centre for Doctoral Training in Cloud Computing for Big Data (EP/L015358/1)}}
}

\name{Georgia Atkinson\textsuperscript{*}, Nick Wright\textsuperscript{*}, A. Stephen McGough\textsuperscript{$\dagger$} and Per Berggren\textsuperscript{$\ddagger$}}
\address{\textsuperscript{*}School of Engineering, \textsuperscript{$\dagger$}School of Computing, \textsuperscript{$\ddagger$}School of Natural \& Environmental Sciences \\
            Newcastle University, United Kingdom \\
            \{g.atkinson, nick.wright, stephen.mcgough, per.berggren\}@ncl.ac.uk}

\begin{document}
\maketitle

\begin{abstract}
    In recent years generative adversarial networks (GANs) have been used to supplement datasets within the field of marine bioacoustics. This is driven by factors such as the cost to collect data, data sparsity and aid preprocessing. One notable challenge with marine bioacoustic data is the low signal-to-noise ratio (SNR) posing difficulty when applying deep learning techniques such as GANs. This work investigates the effect SNR has on the audio-based GAN performance and examines three different evaluation methodologies for GAN performance, yielding interesting results on the effects of SNR on GANs, specifically WaveGAN.
\end{abstract}


\section{Introduction}
    Generative adversarial networks (GANs) are used in a variety of settings within the audio domain, including speech synthesis \cite{BDD19}, music generation \cite{JLY20}, and bioacoustic synthesis \cite{ZHY22}. In marine bioacoustics, GANs have previously been employed for tasks such as dataset generation and augmentation \cite{LLF19,LRK23, YEP22}. Some bioacoustic data, like those from deployed passive acoustic monitoring systems, typically have a low signal-to-noise ratio (SNR). This can cause challenges when generating synthetic data with GANs. 
    
    Despite this, there is little work on the effect of low SNR on GAN-generated synthetic bioacoustics. As such, this work examines how the SNR of marine bioacoustic data affects GAN performance. Further, this work investigates three different methodologies for evaluating GAN performance.

\section{Related Work}


    In 2018, Donahue \etal introduced WaveGAN \cite{DMP18}, an audio-based GAN based on DCGAN \cite{RMC15} adapted to a one-dimensional network. To address the issue of chequerboard artefacts in DCGAN-generated images, WaveGAN phase-shuffles the discriminator at each layer.

    Evaluating GAN-generated bioacoustics typically focuses on improving model performance whilst ignoring the synthetic data's quality \cite{LLF19,LLP20,LRK23}. Work assessing synthetic data production often makes use of Dynamic Time Warping \cite{YEP22}, visually comparing  frequencies to that of similar real-world data \cite{ZHY22}, spectrogram inspection \cite{ZHY22}, cluster analysis \cite{MS21,WWW23}, training discriminator networks to obtain error networks \cite{WWW23}, and Siamese Neural Networks (SNNs) to generate similarity scores between the real and generated data \cite{MS21}.
 
    There is currently no standardised evaluation method for synthetic audio data. Additionally, the effect of SNR on audio generation has not been explored in the marine context.

\section{Data}
     \begin{figure*}[t]
        \centering
        \includegraphics[width=0.95\textwidth]{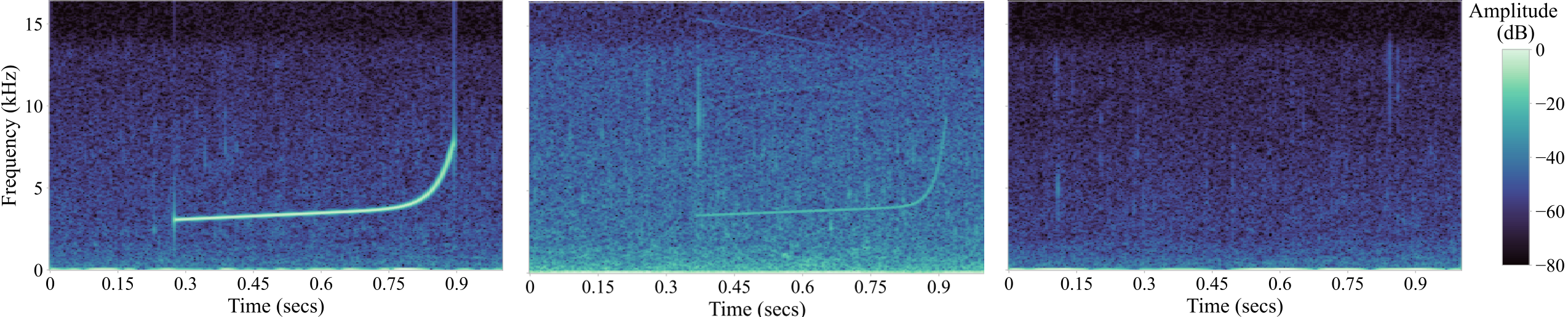}
       \caption{Example of the SNR data. (Left) $\text{SNR} = 10$dB. (Middle) $\text{SNR} = -15$dB. (Right) Background noise.}
        \label{fig:data}
    \end{figure*}
    
    Neural networks which operate on raw audio are often sample rate specific. WaveGAN is designed for audio sampled at 16kHz \cite{DMP18}, making it suitable for marine acoustics at frequencies $\leq$ 8kHz. To evaluate the efficacy of WaveGAN at lower sampling rates, data was generated at 32kHz\footnote{Data generated using: \url{https://github.com/chrisdonahue/wavegan}}.

    Audio of a specific SNR is generated via $\bm{x} = \bm{s} + \beta\bm{n}$, where $\bm{s}$ is the upsweep signal, $\bm{n}$ is the noise signal, $\mathrm{SNR}$ is the SNR, and $\beta = \frac{A_s}{A_n\sqrt{\mathrm{SNR}}}$ is the SNR scaling parameter. $A_j$ is the root mean square amplitude of signal $j$.

    To investigate the effect of SNR on GANs, an upsweep frequency shape which mimics cetacean whistles is used as the signal, chosen as it is a simple shape but one which could reasonably be present within marine bioacoustic data \cite{DL11}. Whistles of varying height, length, and gradient were generated at 32kHz. One-second clips of underwater noise, devoid of anthropogenic activity or whistles, were randomly selected, downsampled to 32kHz, $\beta$-scaled, and combined with the generated whistle. SNR data is systematically generated to cover the range $[-15,10]$dB, at 5k clips per SNR. An additional 5k noise clips are sampled for analysis. See \Cref{fig:data} for example data.

\section{Methodology}

    Diverse synthetic SNR data is generated using a WaveGAN model with a 256-d generator latent space ($\bm{z} \in \mathbb{R}^{256}$ where $z_i \sim U(-1,1)$ for $i=1,...,256$). A model is trained for approximately 10k epochs, chosen empirically based on prior convergence experiments. Each trained network is used to generate 500 synthetic whistles for evaluation purposes.

    To assess synthetic data quality, three methods are explored: frequency spectra comparison; spectrogram pixel intensity comparison; SNN analysis, akin to \cite{MS21}.

\subsection{Frequency Spectra Comparison}\label{subsec:freq_dist}

    Direct signal comparisons, such as Dynamic Time Warping, are unsuitable for comparing GAN-generated signals to training data as they aim to imitate the training data distribution rather than replicate the data itself \cite{GCH14}. To assess the synthetic data generated in this study, first only the frequency domain is considered to provide an overview of each the whilst and noise data.

    Average frequency spectra are calculated to obtain approximate noise and whistle audio samples. Pearson's correlation coefficient is computed for each average spectrum compared to the synthetic sample's frequency spectrum sample, denoted $c_{f,w}$ for whistles and $c_{f,n}$ for noise. If $c_{f,w} \geq c_{f,n}$ the synthetic sample is scored 1 else 0. This is repeated for each synthetic sample, with the mean score calculated for each SNR level.

\subsection{Spectrogram Pixel Intensity Comparison}

    Converting data from the time domain to the frequency-time domain allows for the simultaneous analysis of both domains by examining the pixel intensity distribution in the resulting spectrogram. The similarity between two spectrograms can thus be calculated using pixel intensity distribution correlation. For simplicity, spectrograms used in this work are first converted to greyscale.

    Average pixel intensity distributions (PIDs) of the whistles and noise are calculated to obtain an estimate of their respective distributions. Pearson's correlations between the PID of the synthetic sample and the average PIDs of the corresponding whistle and noise data are calculated, denoted $c_{p,w}$ and $c_{p,n}$ respectively. If $c_{p,w} \geq c_{p,n}$ then the synthetic sample score $ = 1$ else $0$. The overall SNR score is obtained by computing the mean of all synthetic sample scores for that particular SNR, repeated for each SNR. 

\subsection{Triplet Loss Learning via SNNs}

    Two SNN evaluation methods are explored. The first trains a separate SNN for each SNR, utilising the same noise data. The second trains a single SNN on the entire SNR and noise data. Hyperparameters are tuned using Optuna \cite{Optuna} encompassing the number of convolution blocks, dense layers, learning rate and embedding size. Data is split using a 60:20:20 train-test-validation ratio, preprocessed into greyscale spectrograms. SNNs are trained on training and validation data, test data is used for subsequent analysis.  

\subsubsection*{Individual SNNs}
   \begin{figure}[h]
        \centering
		\begin{subfigure}{0.25\textwidth}
			\includegraphics[width=\textwidth]{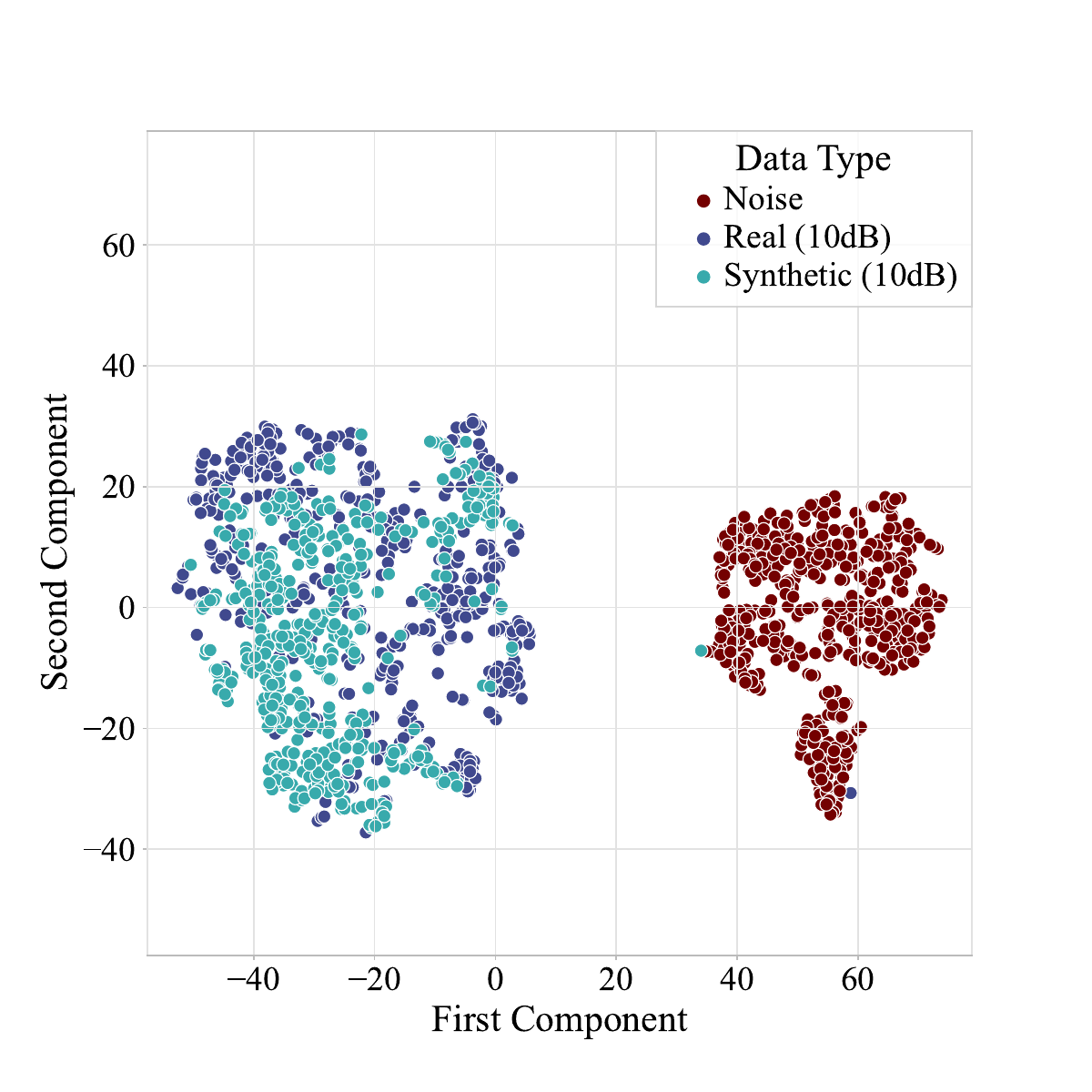}
		\end{subfigure}\hspace{-2em}~
		\begin{subfigure}{0.25\textwidth}
			\includegraphics[width=\textwidth]{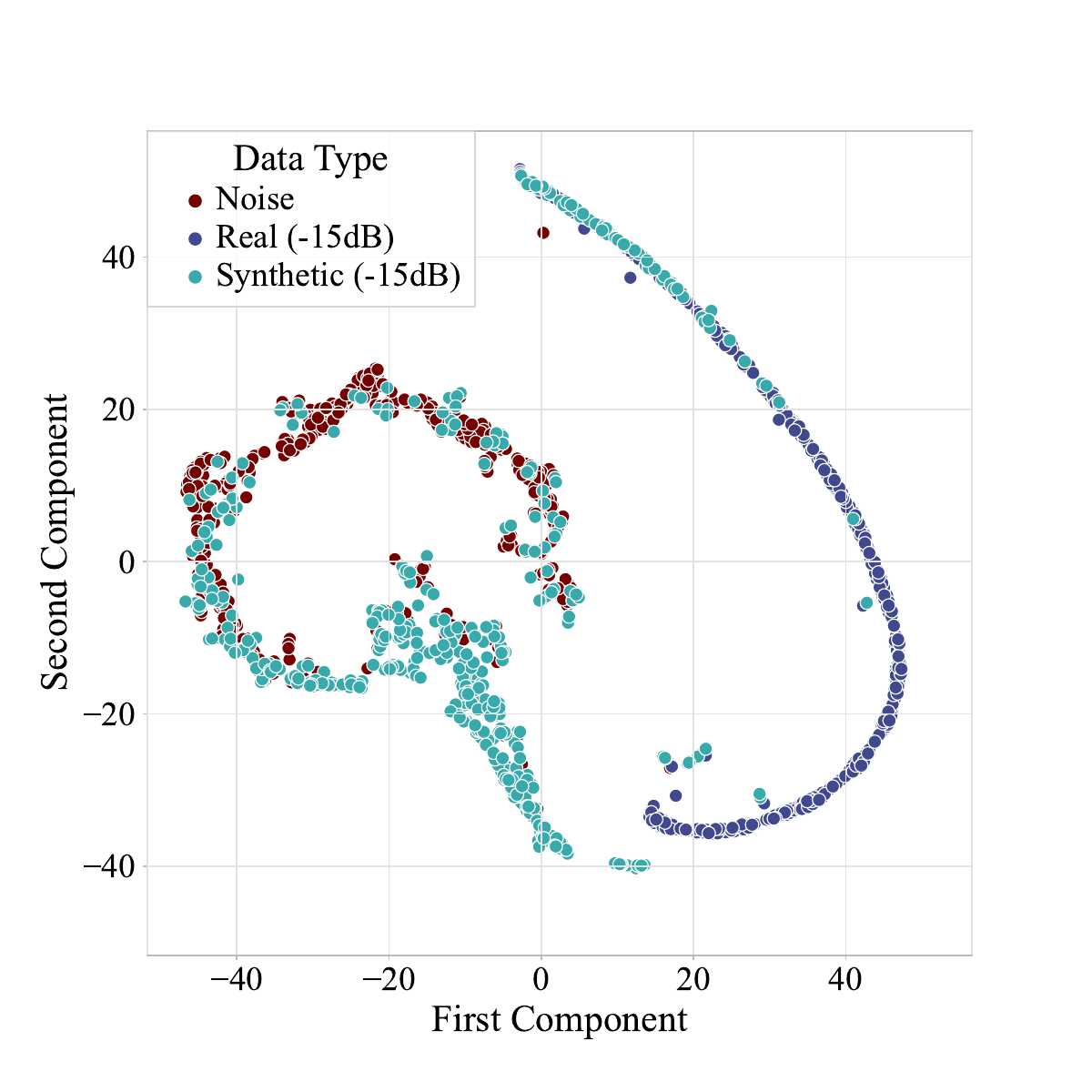}
		\end{subfigure}
        \caption{Example embeddings from the $SNN_{\rm{SNRdB}}$ models visualised using \tsne for the upper and lower SNR bounds.}
  	  \label{fig:snn_some_snr_emb}
    \end{figure}

    \begin{figure*}[t]
        \centering
        \includegraphics[width=\textwidth]{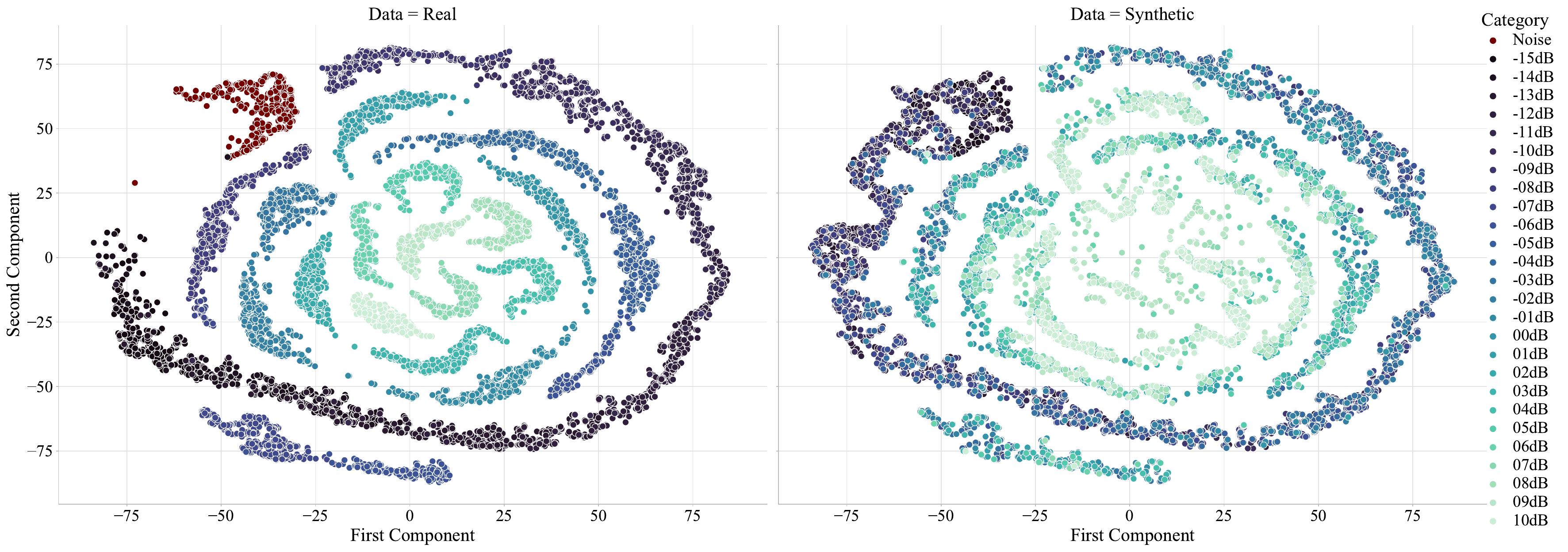}
        \caption{SNN embeddings of real (left) and synthetic (right) data visualised into two dimensions using \tsne. The \tsne visualisations, show the synthetic data is not clustering to the correct SNRs well, with A lot of the points clustered around the noise. It appears that the synthetic embeddings are clustering to a lower SNR decibel.}
        \label{fig:snn_tsne_plot}
    \end{figure*}

    Here, one SNN per SNR ($SNN_{\rm{SNRdB}}$) is trained, each processing a dataset containing 10k spectrograms representing one-second audio files, equally divided between whistles and noise. SNNs are trained using semi-hard triplet loss \cite{SKP15}. After hyperparameter tuning, each model is re-trained to obtain a loss of approximately 0.1, obtained through prior empiric test set evaluation, aiming to prevent overfitting and bias between models.

    Embeddings for 500 synthetic whistles are generated by each $SNN_{\rm{SNRdB}}$ which, along with the whistle and noise embeddings, are used to predict labels for the synthetic data. Visualisations of these embeddings, reduced to two dimensions via \tsne are shown \Cref{fig:snn_some_snr_emb} for the upper and lower SNR bounds. There is a clear separation between the whistle and noise $SNN_{10\rm{dB}}$ embeddings, this is also true for the  with $SNN_{-15\rm{dB}}$ embeddings. The $SNN_{10\rm{dB}}$ synthetic data predominantly cluster with the whistle data, except for one outlier. This outlier exhibits noisy characteristics deviating from typical whistle data. The $SNN_{-15\rm{dB}}$ synthetic embeddings show clustering with both the noise and whistle data. Synthetic data labels are determined using the nearest centroid (NC) algorithm via Euclidean distance measurement, and accuracy scores are generated. 

\subsubsection*{Single SNN}

    Here, a single SNN ($SNN_{\rm{alldBs}}$) is trained on all SNR and noise data, containing 130k spectrograms comprising of 5k noise and 5k whistles per SNR. The network outputs a 56-D embedding and consists of 7 convolutional blocks and 2 fully connected layers, trained using the Adam \cite{KB14} optimiser with a learning rate of $2\times10^{-4}$. The network achieves a validation loss of 0.37. 

    In \Cref{fig:snn_tsne_plot}, \tsne plots of the real and synthetic embeddings are displayed. $SNN_{\rm{alldBs}}$ clusters data effectively with higher SNRs centralised and lower SNRs spiralling outwards and distinctly clustered noise data, providing confidence in lower SNR results. Interestingly, synthetic data appears to cluster with lower SNRs relative to their actual SNR, e.g. 10dB synthetic data are clustered around the 3dB or 4dB whistle data. Lower synthetic SNR data often cluster around the noise data or between the $-15$dB whistle and noise data. This is unlike results in \Cref{fig:snn_some_snr_emb}, where synthetic data typically clusters well with the corresponding whistle data. This suggests treating SNRs as continuous rather than categorical motivating classification and regression approaches. 

    The challenge in treating the task as a regression problem lies in handling background noise data. This data can be viewed as an infinitesimally small SNR, effectively $SNR_{dB}\rightarrow -\infty$ or equivalently $SNR \rightarrow 0$, or as a categorical variable, `noise'. In the previous metrics, the latter case is applied since only one decibel level was considered at a time.
    
    In the categorical approach, NC is utilised to predict decibel and noise labels for the synthetic data. Initially, centroids are computed for each SNR and noise data. Subsequently, the synthetic embeddings are passed to the algorithm and Euclidean distance is employed to determine the predicted labels. Conversely, in the regression approach, $K$-nearest neighbours (KNN) is adopted. Instead of an accuracy, root mean square error (RMSE) is employed for evaluation. The optimal $K$ value is determined by employing the elbow method on the RMSE, $R^2$, and the maximum distance between a point and its neighbours from fitting KNN to test data embeddings for $K \in [1,1000]$.

\section{Results}
\subsection{Frequency Spectra Comparison}\label{subsec:r_freq_dist}

\begin{figure}[h]
		\centering
		\includegraphics[width=\linewidth]{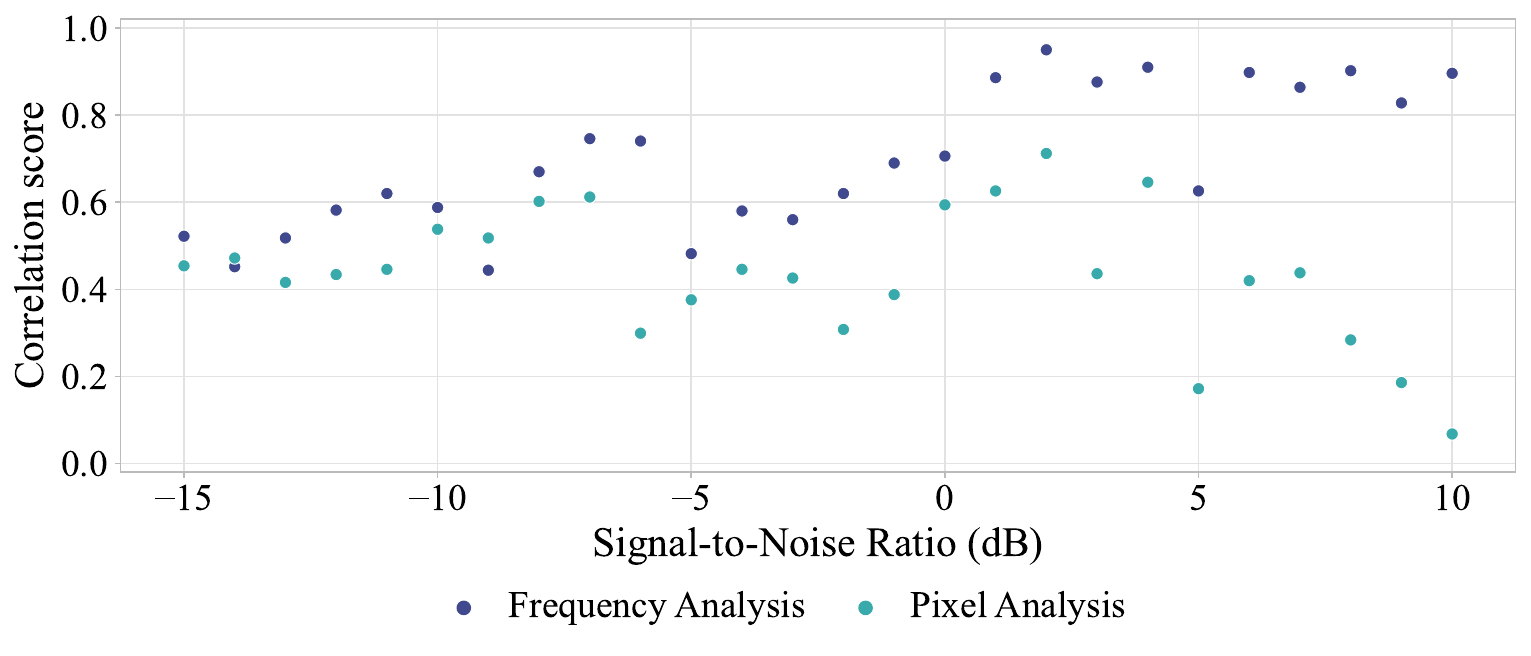}
		\caption{The scaled frequency and pixel analysis correlation scores between real data and synthetic data.}
    \label{fig:freq_and_spec_dist_results}
	\end{figure}

    \Cref{fig:freq_and_spec_dist_results} presents the results of the frequency spectrum analysis. The correlation score increases with the SNR. The increase in correlation score has an approximately linear trend up to 0dB. Beyond this point, within the range $(0, 10]$dB, the correlation scores stabilise around 0.9, except the outlier score observed at 5dB.

\subsection{Spectrogram Pixel Intensity Comparison}

    \Cref{fig:freq_and_spec_dist_results} presents the results of the spectrogram pixel intensity analysis. The correlation score does not exhibit any well-defined trend but primarily decreases within $[5, 10]$dB. As pixel intensity is based upon the gradient of greyscale, for high SNR there is a large number of white pixels compared to the noise data due to the prominent whistle in the audio clip. However, the synthetic samples do not exhibit the same prominent white pixels, which influences the correlation score. This discrepancy may arise from differences in SNR between the synthetic whistles and the whistle data.

\subsection{Triplet Loss Learning}

\subsubsection*{Individual SNNs}
\begin{figure}[h]
		\centering
		\includegraphics[width=\linewidth]{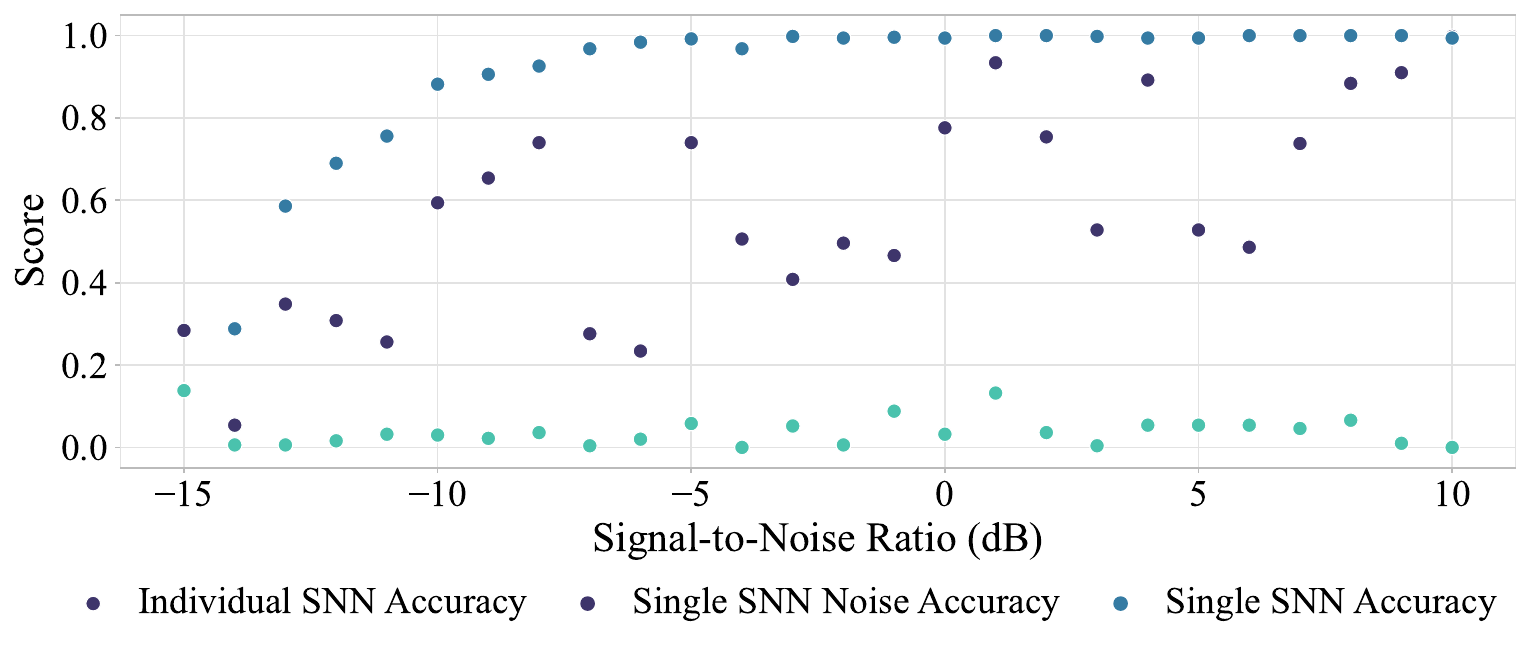}
		\caption{The NC accuracy scores for each SNR obtained via individual SNNs, single SNN with the classic accuracy and noise label accuracy.}
		\label{fig:snn_all_data_nc}
	\end{figure}

	\Cref{fig:snn_all_data_nc} displays results for the individual SNNs (along with other results discussed later). This reveals an approximately linear relationship between the SNR and accuracy score, as in \Cref{subsec:r_freq_dist}.  Notably, this method heavily relies on ensuring consistent performance across 26 distinct SNN models without overfitting is challenging. The obtained losses in $SNN_{05dB}$ and $SNN_{06dB}$, are very small (see \Cref{fig:snn_all_data_nc}), deviating from the general trend suggesting potential overfitting. 
	
\subsubsection*{Single SNN}

	\Cref{fig:snn_all_data_nc} presents the results obtained via categorical analysis. Assessing network accuracy based on whether the correct SNR decibel is predicted is unsuitable in this context as it does not appropriately capture the relationship between WaveGAN performance and SNR. When focusing on a binary outcome for each SNR, either noise or not (denoted `Single SNN Noise Accuracy'), the results indicate WaveGAN models produce synthetic data resembling whistles consistently from -5dB onwards. However, this approach tends to penalise lower SNRs.
	
    This ambiguity in evaluating data in a categorical manner prompted a shift to considering SNR labels as continuous variables, assigning noise an SNR of 0. Here, SNRs are converted from decibels to the original SNR (e.g. -10dB = 0.1) and the $K$-Nearest Neighbours (KNN) algorithm is applied to the test embeddings. SNRs are thus predicted using both the mean and weighted mean of a point's $K$ neighbours, scaled based on the distance to the point. 
    
    Through the elbow method, the optimal $K = 400$ is determined via weighted mean. Synthetic embeddings are passed to KNN to obtain an SNR prediction value, then converted back to dB to ensure a fair comparison between SNRs. For example, both -14dB to -15dB and 9dB to 10dB have a difference of 1dB, though when considered a ratio the difference is 0.0082 and 2.06 respectively, thus penalising higher SNRs. The minimum non-zero SNR predicted is $-37.5\text{dB}$, hence SNRs predicted at 0 are assigned a value of $-40\text{dB}$, ensuring only lower SNRs are penalised. The root mean square decibel error (RMSDE) is then calculated using the difference between predicted and actual SNR in dB.

    \begin{figure}[ht]
		\centering
		\includegraphics[width=\linewidth]{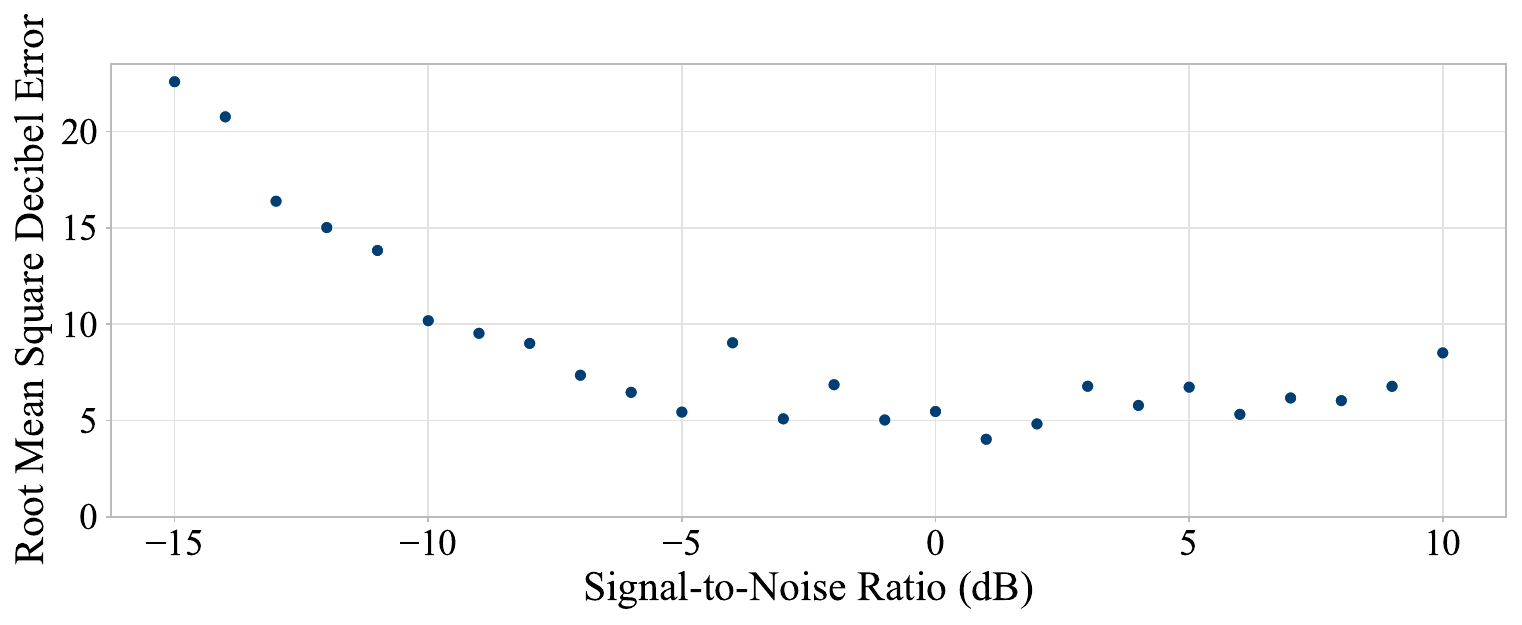}
		\caption{The RMSDE of the synthetic embeddings. The RMSDE reaches a minimum value in the range $[-2, 2]$dB.}
		\label{fig:snn_all_data_knn_rmse}
	\end{figure}

    \Cref{fig:snn_all_data_knn_rmse} shows the RMSDE for each SNR. Most SNRs within [-10,10]dB have an RMSDE within [5,10], indicating the average distance between the synthetic data's actual and predicted SNR is [5,10]dB. Notably, lower SNRs perform less satisfactorily. Unlike other approaches, this evaluation takes into account the WaveGAN generator's ability to produce audio with the correct SNR. Evaluating test embeddings using this metric, all SNRs have an RMSDE within [0,0.8], enhancing confidence in the results shown. These findings reinforce the observation that synthetic data tends to have a lower SNR compared to the target SNR for WaveGAN models, amplified for lower SNRs. 
    
\section{Conclusion}

    Comparing frequency spectra provides a simple method for evaluating WaveGAN's general performance at an individual SNR level. However, this is limited for lower SNRs due to the similarities in frequency spectra and marginal differences in correlations. Additionally, it does not capture whether a correct sound is created. Comparing this to pixel intensity evaluation, this method particularly struggles due to the large amount of dark and white colouration within the low and high SNR spectrograms with an inability to capture whistle-like characteristics. Unlike these methods the SNN methods exhibit a clear distinction between the noise and whistle data for lower SNR, giving higher result confidence.

    Investigation using SNNs suggests poor evaluation may be the result of WaveGAN producing audio at a lower SNR relative to that with which it is trained. This study has shown SNNs can be applied to evaluate individual SNR performance and synthetic data quality, as well as to gauge if the correct SNR is generated by the WaveGAN by assessing all SNRs collectively, generating an easily interpretable RMSDE metric which indicates WaveGAN's proficiency for generating synthetic data at the correct SNR.  
    
    Further, this work demonstrates the value of utilising SNNs for audio generation evaluation. SNNs do not require large data volumes, making them suitable for assessing generated data in scenarios with limited real-life examples. Based on the results obtained, it is recommended to train one SNN over all SNR classes if using multi-class data, including one class for noise if possible. Visual inspection using \tsne is recommended to assess synthetic data spread. 
        
    For all evaluation methods, it is evident that SNR significantly impacts WaveGAN performance, with lower SNRs resulting in poorer performance. Notably, WaveGAN on average produces a lower SNR than the training data, as demonstrated in \Cref{fig:snn_tsne_plot}. As such, it is recommended to train WaveGAN models on data with an $\text{SNR} \geq -5\text{dB}$.

    Future work should explore the impact of SNR on other audio-based GANs, such as WaveNet \cite{ODZ16}, DiffWave \cite{KPH20}, or MelGAN \cite{KKD19} following a similar analysis. Additionally, it is worth considering alternative similarity metrics, such as Jaccard similarity and Cosine similarity. Rather than assigning a binary score of 1 or 0 based on correlation values, the raw correlation values between the whistle data and synthetic samples could be utilised. Finally whilst this work located a clear lower performance limit of $-15\text{dB}$, future work should explore the existence of an upper limit. 

{
\balance
\bibliographystyle{IEEEtran}
\bibliography{main}
}

\end{document}